\documentclass{IEEEcsmag}

\usepackage[colorlinks,urlcolor=blue,linkcolor=blue,citecolor=blue]{hyperref}
\expandafter\def\expandafter\UrlBreaks\expandafter{\UrlBreaks\do\/\do\*\do\-\do\~\do\'\do\"\do\-}

\usepackage{url}
\usepackage{amsmath}
\usepackage{upmath,color}
\usepackage{siunitx}

\jvol{XX}
\jnum{XX}
\paper{8}
\jmonth{Month}
\jname{Publication Name}
\jtitle{Publication Title}
\pubyear{2021}

\begin{document}

\sptitle{THEME ARTICLE: Energy}

\title{Wireless charging and readout via textile coil for continuous full-body wearable computing}

\author{Ryo Takahashi}
\affil{The University of Tokyo, Tokyo, 1138656, Japan}

\author{Yoshihiro Kawahara}
\affil{The University of Tokyo, Tokyo, 1138656, Japan}

\markboth{THEME/FEATURE/DEPARTMENT}{THEME/FEATURE/DEPARTMENT}

\begin{abstract}
The growing use of wearable devices for activity tracking, healthcare, and haptics faces challenges due to the bulkiness and short lifespan of batteries. Integration of a textile-based wireless charging and readout system into everyday clothing can enable seamless power supply and data collection around the body. However, expanding such system to cover the entire body is challenging, as it increases electromagnetic interference with the body, degrading the performance of wireless system. This article introduces a meandered textile coil designed for body-scale, efficient wireless charging and readout.  The meander coil can confine a strong inductive field near the body surface, ensuring W-class safe charging and sensitive readout with uW-class low power. Moreover, its zigzag design is simple enough for mass production on industrial knitting machines. Therefore, the body-scale meander coil can continuously operate battery-free wearable devices across the body, leading to --\textit{Internet of Textiles}-- ubiquitous deployment of continuous full-body wearable computing into everyday clothing.
\end{abstract}

\maketitle

\begin{table*}[t!]
 \centering
 \caption{Application examples of wearable computing, including activity logging, healthcare monitoring, haptic feedback, and augmented/virtual reality.  * denotes a handheld controller associated with a head-mounted display.}
  \includegraphics[width=2.0\columnwidth]{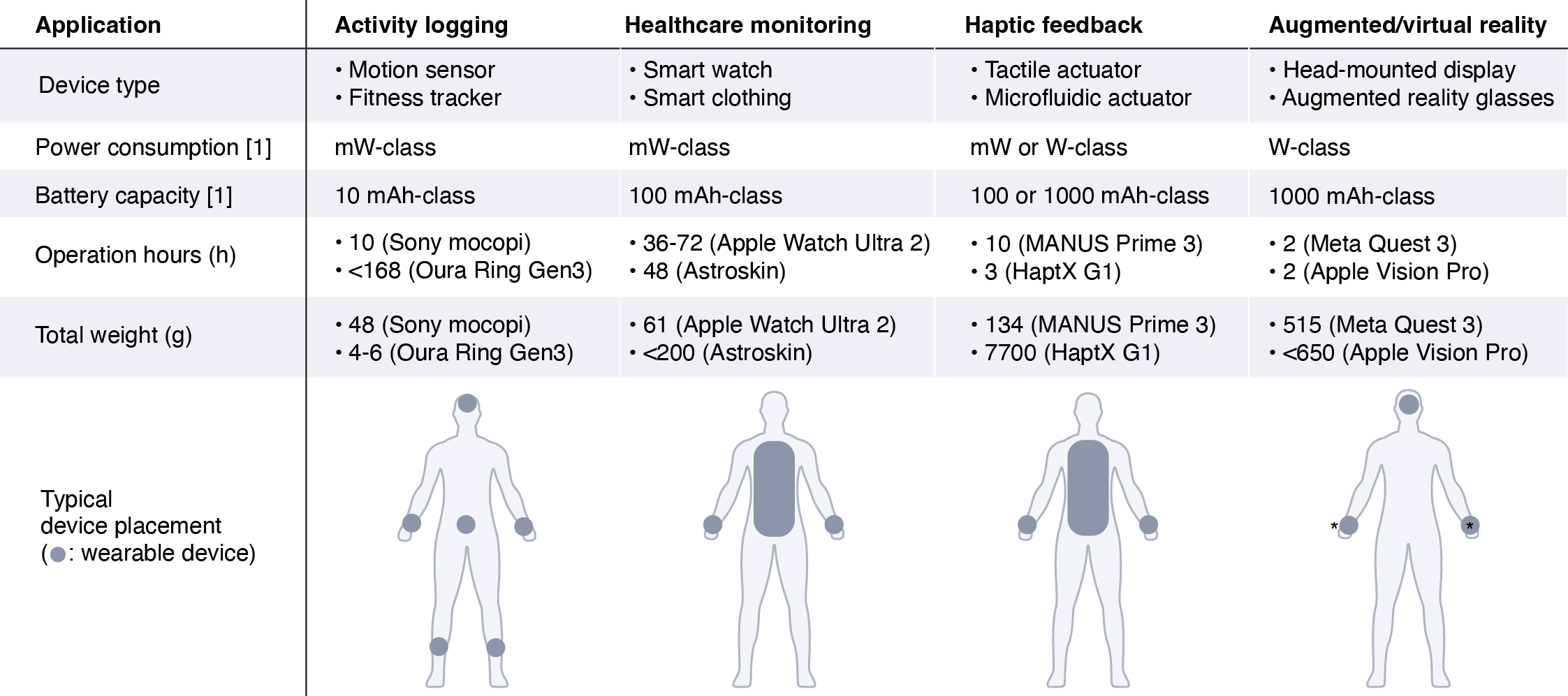}
\label{tab:wearable}
\end{table*}

\begin{figure*}[t!]
 \centering
  \includegraphics[width=2.0\columnwidth]{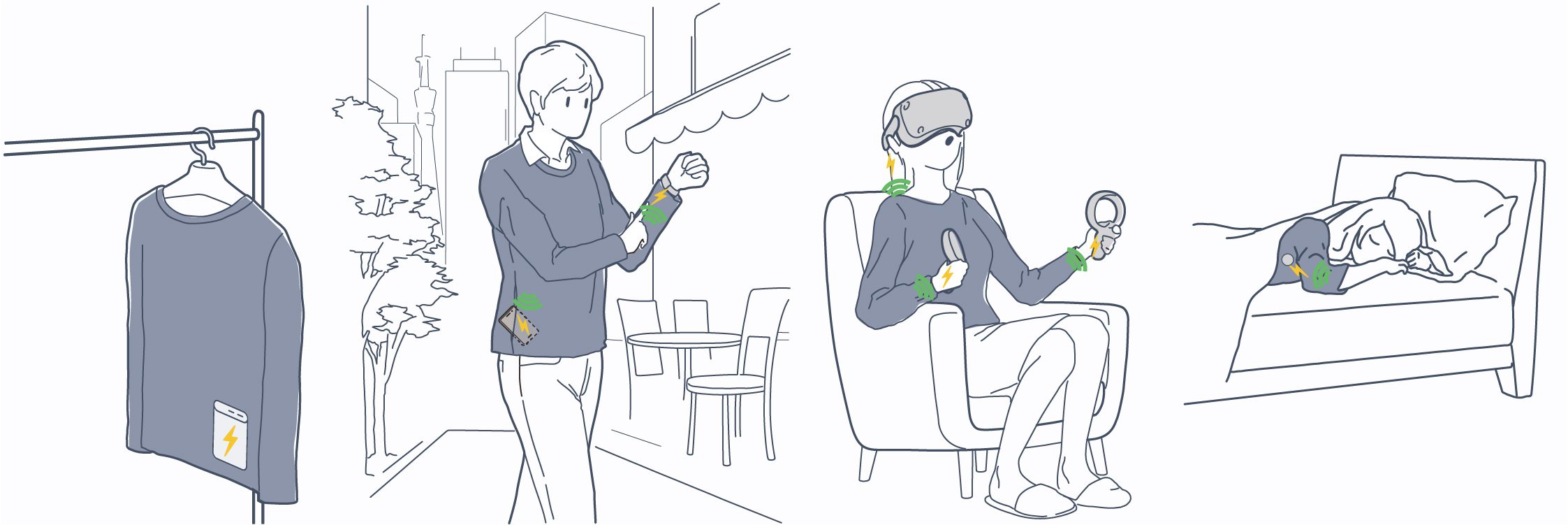}
\caption{Illustration of textile-based wireless charging and readout. A gray-colored electric textile connected to a mobile battery can construct wireless power transfer and data collection scheme for on-body wearable devices throughout the textile. Such clothes has the potential to allow a battery-free, lightweight design of wearable device operating onto the body for days and weeks.}
\label{fig:vision}
\end{figure*}

\chapteri{T}he development of electronics technology has led to the dramatic miniaturization and acceleration of computer systems, fundamentally altering our interaction with digital technology. 
This shift has lead to the rise of wearable computing, seamlessly merging computing functions with our daily clothes and accessories. 
Wearable computing can facilitate extensive interaction with the digital realm, ensuring constant connectivity even during passive engagement with technology. 
Nevertheless, for wearable computing to realize its full potential, devices need to be not only compact but also capable of continuous operation on the body.

Yet, conventional wearable devices are typically powered by batteries, despite the miniaturization of electronics for sensing, actuation, and communication modules.
The bulkiness and limited lifespans of batteries compromise user comfort and necessitate frequent charging~\cite{Sun2023Requirements}.  
\autoref{tab:wearable} lists the technical specifications of four application examples of wearable computing: activity logging, healthcare monitoring, haptic feedback, and augmented/virtual reality~(AR/VR).
Basically, high-powered, \si{\W}-class wearable devices such as haptic actuators and head-mounted displays, struggle with both short operational time and discomfort of wearing the heavy devices on the head and hands.
On the other hand, low-powered, \si{\mW}-class wearable computing consisting of various types of tiny sensors can be relatively lightweight and operate for extended periods, thanks to the low power consumption of the sensors.
Unfortunately, their built-in battery occupies a substantial amount of the sensor's overall weight and space, posing a challenge to design a thin and compact sensor like an always-attachable band-aid.

Incorporating power supply and communication hubs directly into clothing offers a viable solution for facilitating the continuous operation of wearable devices.
This setup enables battery-free and lightweight design of wearable devices, drawing operational power from external sources while communicating signals to central hubs. 
Direct wiring is a popular approach, yet this might impose physical constraints on both users and devices, as the tethers are tightly connected to the devices~\cite{Wicaksono2020Tailored}.
While conductive fabric allows for flexible connection throughout the textile, safety concern is still unsolved due to the exposed electrodes for electrical connections~\cite{Noda2019I2We,plug-n-play-eknit2025}.
On the other hand, wireless connections, particularly those based on textile-based wireless charging and readout systems, have gained attentions for their ability to naturally attach wearable devices to the body without physical constraints~(see \autoref{fig:vision}).

\begin{table*}[t!]
 \centering
 \caption{Technical comparison of two-dimentional wireless charging and readout techniques around the body.}
  \includegraphics[width=2.0\columnwidth]{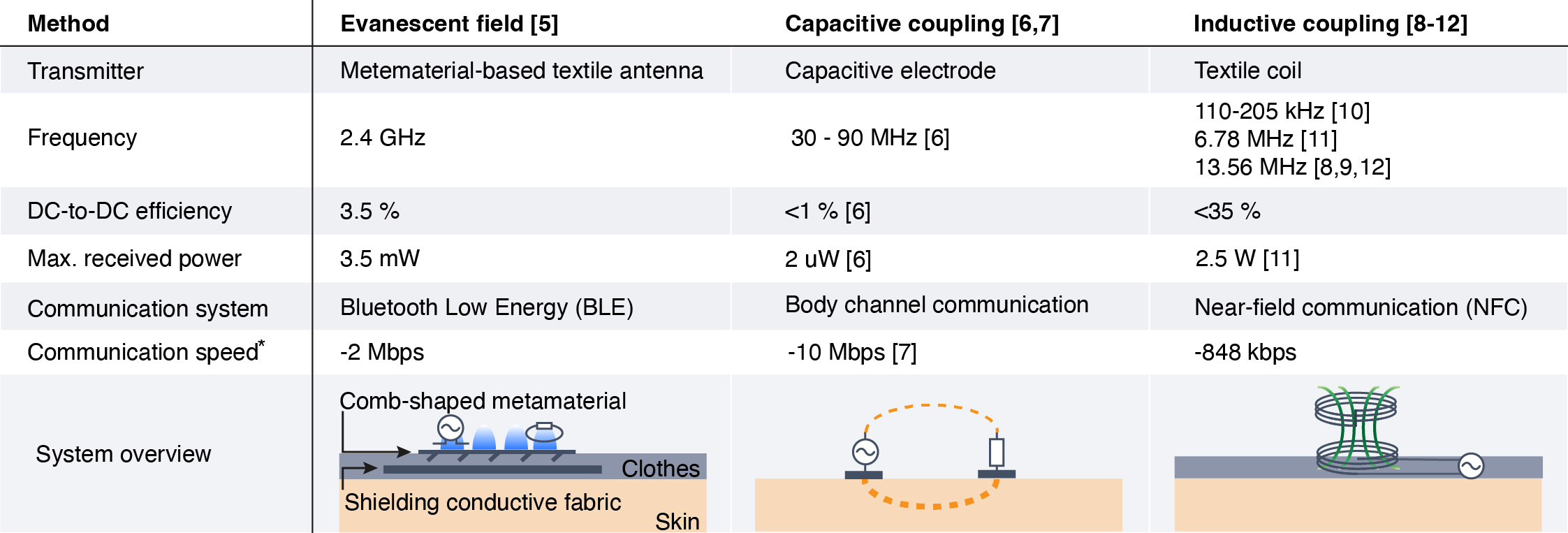}
\label{tab:method}
\end{table*}

Prior research has explored various methods for on-body wireless charging and data readout.
This paper focuses on two-dimensional~(2-D) wireless interconnection approach~(see \autoref{tab:method}), which restricts the area of wireless charging and readout to near the skin, by using evanescent field~\cite{Tian2019Metamaterial}, capacitive coupling~\cite{Li2021BodyPower, Lee2020BodyComm}, or inductive coupling~\cite{Lin2020TextileCoil, Lin2022GainCoil, Li2018PrintCoil, Takahashi2022MeanderCoil++, Takahashi2022TwinMeadnerCoil}.
Compared to standard 3-D wireless connections like Wi-Fi and Bluetooth Low Energy~(BLE), 2-D wireless approach can offer energy-efficient and secure connection with the wearable devices onto the body, while suppressing the radiation of wireless signal.
Because the wearable devices are typically placed near the body surface, such 2-D wireless networking is compatible with the on-body connection of wearable devices.
Among the various approaches, coil-based inductive coupling can efficiently charge and read out wearable devices~\cite{Lin2020TextileCoil, Lin2022GainCoil, Li2018PrintCoil, Takahashi2022MeanderCoil++, Takahashi2022TwinMeadnerCoil}.
This is because the inductive coupling can interact less with the dielectric body, compared to the other approaches such as evanescent field~\cite{Tian2019Metamaterial} or capacitive coupling~\cite{Li2021BodyPower, Lee2020BodyComm}.
Thereby, this article focuses on a coil-based wireless charging and readout across body toward continuous power delivery and communication system for wearable devices.

\section{OVERVIEW OF MEANDER COIL}
\label{sec:overview}

\begin{figure*}[t!]
 \centering
  \includegraphics[width=2.0\columnwidth]{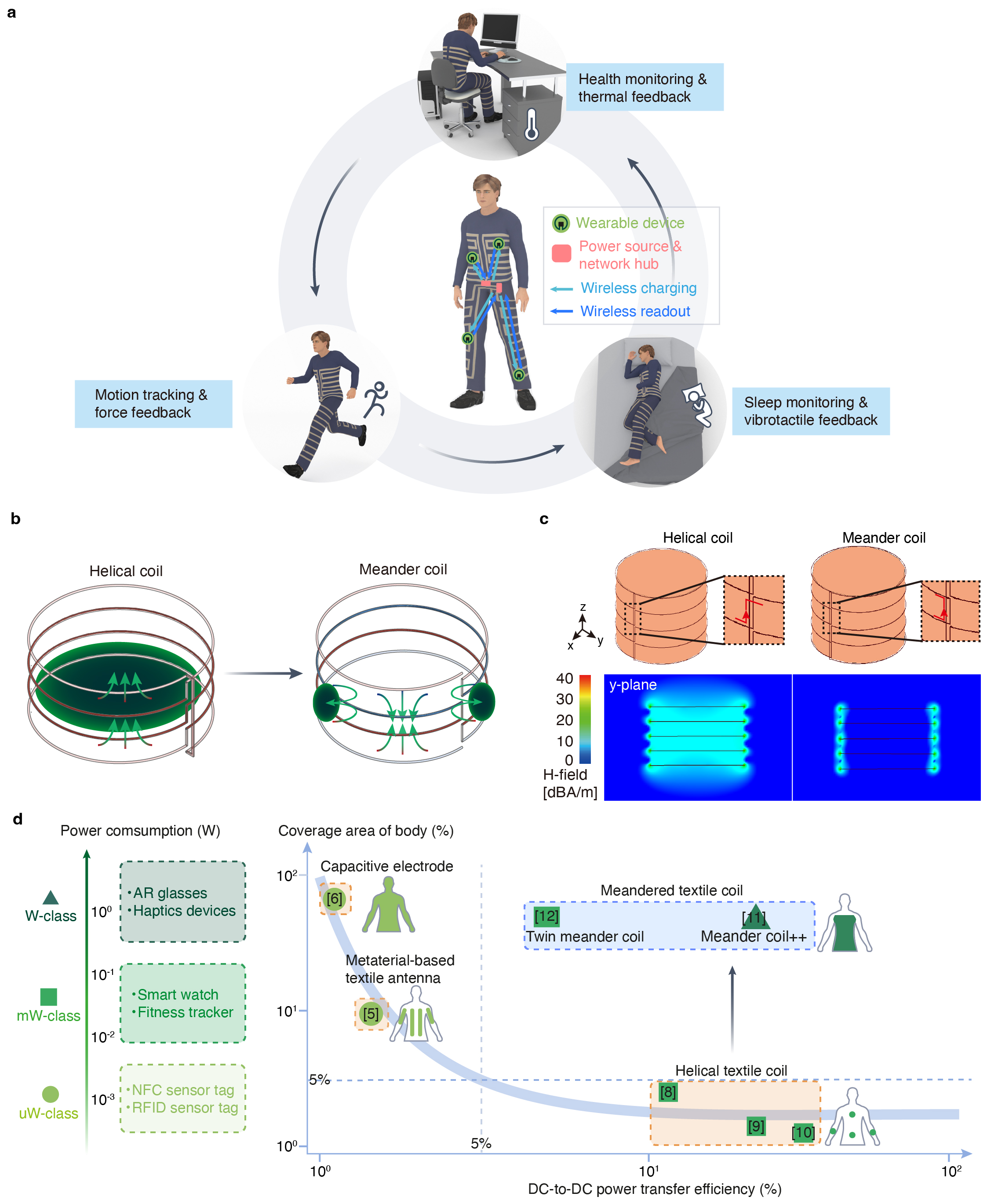}
  \caption{Overview of wireless charging and readout using a body-scale meandered textile coil. (a) Illustration of meandered textile coil. (a) Illustration of wiring structure of helical or meander coil and (b) their simulated inductive field pattern near the simplified circular body model. (c) Technical comparison of the meandered textile coil with prior approaches in terms of DC-to-DC power transfer efficiency and coverage area of the body.}
  \label{fig:overview}
\end{figure*}

Despite an energy-efficient advantage of the inductive approach, the development of a large-scale textile coil presents significant challenges.
First, a standard helical coil close to the body causes an electromagnetic~(EM) interaction with the body, because the helical coil generates a strong inductive field in the body proportional to the coil size.
Such an EM interference weakens the inductive coupling, leading to significant losses in wireless transfer efficiency and wireless readout performance.
Moreover, multi-layered wiring pattern of the standard coils constrains the fabrication process to the post-process techniques like digital sewing~\cite{Lin2020TextileCoil, Lin2022GainCoil} and screen printing~\cite{Li2018PrintCoil}, restricting the fabrication of textile coil on a small scale.   
As a result, prior textile coils are limited to only a few \% coverage of the entire body to achieve stable power and communication link via a textile coil~\cite{Lin2020TextileCoil, Lin2022GainCoil, Li2018PrintCoil}.
These narrow-range operation and small-scale fabrication pose fundamental limitations to the placement of the textile coil, requiring cumbersome and specialized design for each specific wearable application.

To solve this challenge, we previously introduced a meander coil that enables wireless charging and readout of the wearable devices placed anywhere~(see \autoref{fig:overview}a)~\cite{Takahashi2022MeanderCoil++, Takahashi2022TwinMeadnerCoil}.
Unlike the helical coil which wraps the wiring in the same direction, the meander coil which alternates the winding direction between clockwise and counter-clockwise for one or few turns.
This design offers two main advantages over the traditional helical coil structure.
First, the meander coil can confine strong inductive field the body surface while cancelling the penetration of the inductive field inside the body~(see \autoref{fig:overview}b).
The result of EM simulation for both a helical and meander coil shows the confinement of the strong inductive field within \SI{1}{\cm} of the coil edge while significantly reducing the strength of the inductive field by approximately \SI{10}{\dB} over $90\%$ of the entire body~(see \autoref{fig:overview}c).
This allows the body-scale meander coil to improve wireless power performance and wireless readout sensitivity, as will be detailed later.
Secondly, the meander coil can be easily fabricated using industrial machine knitting~\cite{Takahashi2022TwinMeadnerCoil}, because the meandered wiring pattern is one-sided simple zigzag pattern.
This can provide the strong possibility for mass production of the textile coil.

Current wearable devices are divided into three groups based on their power consumption:~\si{\W}-class devices such as AR glass and haptic actuators, \si{\mW}-class devices such as smart watch and fitness tracker, and \si{\uW}-class devices such as near-field communication~(NFC) and radio-frequency identification~(RFID) sensor tag.
Referring to this category, we compare the meandered textile coil with prior three researches including metamaterial-based textile antenna, capacitive electrode, and helical textile coil, based on power transfer efficiency and coverage area of the body, as well as the support level among three groups of the wearable devices~(see \autoref{fig:overview}d).
Basically, prior approaches suffer from either low power transfer efficiency~($<5\%$) or narrow coverage area~($<5\%$), owing to the EM interaction with the body.
In contrast, the meandered textile coil can achieve both large-area, energy-efficient, \si{\W}-class, wireless charging across the body thanks to its localized inductive pattern.

\section{EXAMPLES OF MEANDER COIL}
\label{sec:example}

\begin{figure*}[t!]
 \centering
 \includegraphics[width=2.0\columnwidth]{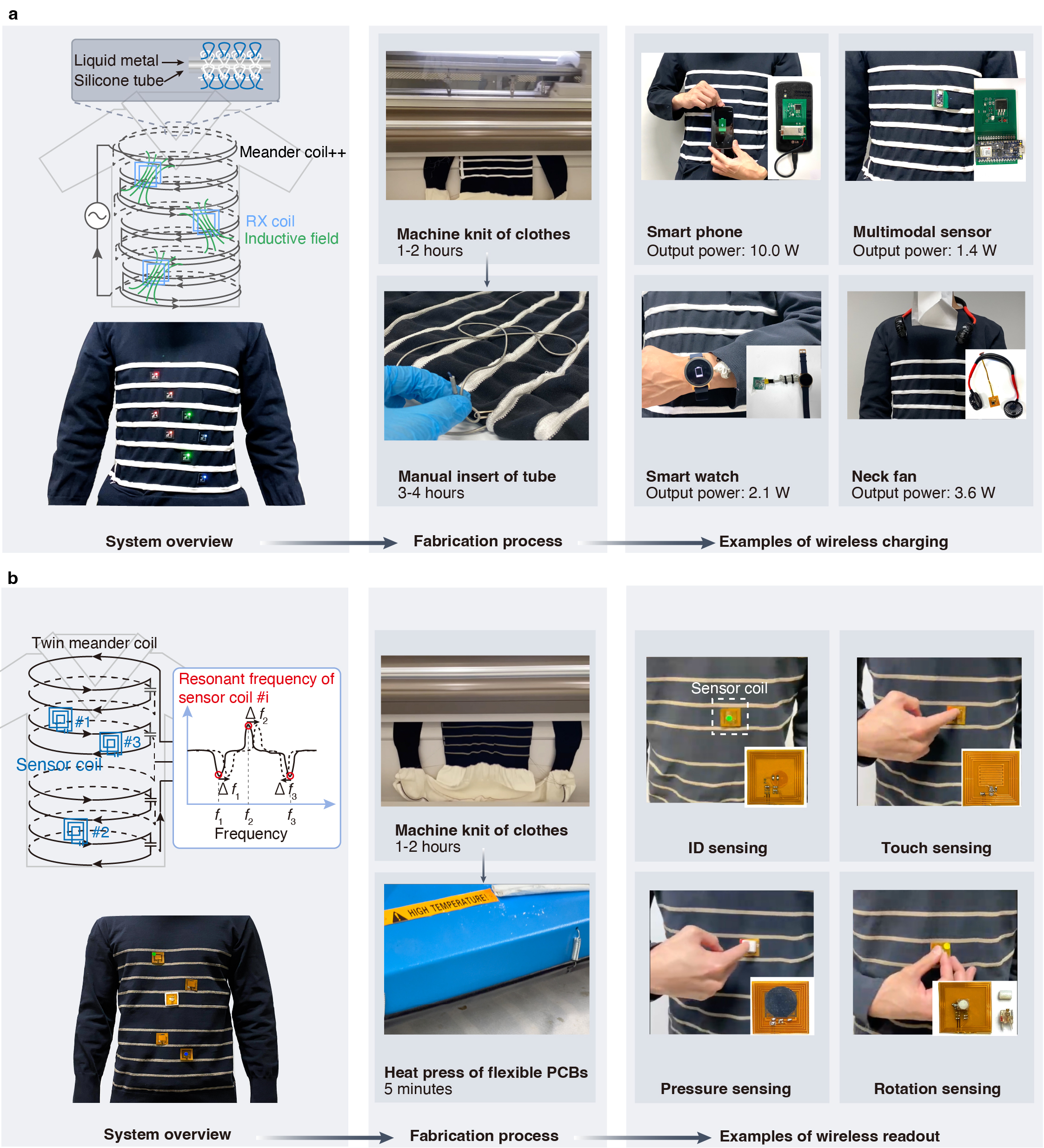}
 \caption{Examples of meandered textile coil. (a) Liquid-metal-based meander coil named meander coil++ can safely and efficiently send \si{\W}-class power to the wearable devices. The output power indicates the transferred power from the meander coil++. (b) Machine-knitted meander coil named twin meander coil can provide a fast and automatic fabrication of the meandered textile coil with a minimal post processing, in addition to supporting \si{\uW}-class wireless readout of battery-free wearable sensors. The wearable sensors include four sensor coils, each of which can superimpose ID, touch, discrete rotation, or pressure value into a unique passive response, have different resonant frequencies for the twin meander coil to recognize the sensor's type.}
 \label{fig:example}
\end{figure*}

This section explains two examples of meandered textile coil.
By combining the meander coil with a liquid-metal-based low-loss conductive
thread~\cite{Takahashi2022MeanderCoil++} or a sensitive readout system~\cite{Takahashi2022TwinMeadnerCoil}, the body-scale meander coil can achieve \SI{2.5}{\W} safe charging at $25\%$ DC-to-DC efficiency~($41\%$  AC-to-AC efficiency), or \si{\uW}-class low-powered readout of a battery-free wearable device occupying less than $0.5\%$ of the coverage area.

\subsection{Liquid-metal-based meander coil}

First, we introduce meander coil++~\cite{Takahashi2022MeanderCoil++,sato2025mems}, which combines the meander coil with a low-loss conductive thread based on liquid metal~(see \autoref{fig:example}a).
For an energy-efficient and body-scale textile coil, a stretchable, low-loss, conductive thread is preferred, yet, the usual stretchable conductive threads are so resistive.
To solve this challenge, we employ eutectic gallium indium (eGaIn) enclosed within a stretchable silicone tube as a flexible and low-energy-loss thread.
eGaIn has high conductivity and minimal toxicity, and it also remains in a liquid state at room temperature~\cite{Dickey2008eGaIn}.
The resistivity of the liquid-metal-based conductive thread with a \SI{1}{\mm} diameter~(\num{29.6e-8}~$\Omega$m) is about ten times lower than that of the standard conductive thread~(\num{246.6e-8}~$\Omega$m)~\cite{Takahashi2022MeanderCoil++}.
As a result, the body-scale meander coil++ can efficiently transmit \si{\W}-class power to the small wearable devices around the body, allowing continuous operation of the battery-free wearable devices.
Furthermore, the body-scale wireless charging provides the flexibility for the different arrangement of the wearable devices according to application scenarios, without the need to change the placement of textile coil.

The meander coil++ can be fabricated as follows: first, a knitting machine called Whole Garment~(MACH2XS 15S, Shima Seiki) automatically produces a clothing with specially designed pockets, then users manually insert the conductive thread in the pockets.
The clothing and pocket are composed of stretchable threads (BabyFit, Asahi Kasei Advance Corporation) and a durable thread (Tsunooga, TOYOBO), respectively.
Thanks to the use of the durable thread, the conductive thread in meander coil++ is protected against daily wear and wash.
The measured AC-to-AC and DC-to-DC efficiency at \SI{6.78}{\MHz} from meander coil++ to a \numproduct{4x4}~\si{\cm} receiver coil occupying $0.4\%$ of meander coil++ were approximately $41\%$ and $25\%$, respectively~\cite{Takahashi2022MeanderCoil++}.
Such relatively high efficiency allows meander coil++ to safely transmit \si{\W}-class power to the wearable devices, including a LED display, a smartphone, a smart watch, an Arduino-based multi-modal sensor, and a neck fan.
Using an EM simulator~(FEKO, Altair), we also confirmed that meander coil++ can transmit up to \SI{33}{\W} power~\cite{Takahashi2022MeanderCoil++}, while complying with international safety guidelines regarding electromagnetic exposure~\cite{ICNIRP2020}.
Although current meander coil++ relies on relatively power-consuming BLE for readout protocol with the wearable devices, we confirmed a low-powered NFC protocol, which supports low-powered communication between coil-based NFC reader and sensor tag, could be available with the meander coil++, by adjusting its resonant frequency to the NFC operating frequency of \SI{13.56}{\MHz}.

\subsection{Machine-knitted meander coil}

Then, we explain twin meander coil~\cite{Takahashi2022TwinMeadnerCoil}, a coil structure that allows for automatic fabrication of the meander coil through machine knitting of the conductive thread~(see \autoref{fig:example}b).
Although the meander coil++ manually inserts the conductive thread in the pockets to create the meander pattern, the twin meander coil can automatically fabricate the meander pattern.
The only post-process is a quick heat-press to connect the chip capacitors mounted on the flexible PCBs with the textile.
This automation reduces the total fabrication time from approximately $6$ hours to about $2$ hours, offering a fast, scalable fabrication of the meander coil.
However, the power transfer efficiency of twin meander coil becomes low~(about $8\%$) due to the high resistance of the conductive thread~(AGposs, Mitsufuji Corporation) used for our machine knitting.
Although such low efficiency needs low-powered wireless charging and readout like NFC, we examined the output power over \numrange{100}{200}~\si{\mW} is necessary to activate communication modules in an NFC sensor tag.

To lower the output power, the twin meander coil employs a wireless readout technique called passive inductive telemetry~(PIT)~\cite{Takahashi2020TelemetRing, picoRing2024takahashi}, which is capable of gathering the sensor's data from wearable devices with \si{\uW}-class output power.
Generally, PIT consists of an external reader coil and a fully-passive sensor coil.
First, the sensor coil passively changes its resonant frequency according to the value change of passive sensors like capacitive pressure, temperature, or humidity sensors.
When the reader coil inductively couples with the sensor coil, the reader coil can then detect a sharp peak around a sensor's resonant frequency through the frequency response.
The resonant frequency~($f$) is determined by an inductance~($L$) and a capacitance~($C$) of the coil as follows: $f=1/\left(2\pi\sqrt{LC}\right)$.
Based on this formula, finally, the reader coil can estimate the change of the sensor value via the peak shift.
Because the reader coil just emits a weak inductive field to couple with the sensor coil, the reader coil can lower the output power up to \si{\uW}-class.
Moreover, the sensor coil can be battery-free as the sensor coil operates by modifying the inductive field in a passive manner.

\begin{table*}[t!]
 \centering
 \caption{Technical comparison of two types of our meandered textile coils with prior helical textile coils.}
  \includegraphics[width=2.0\columnwidth]{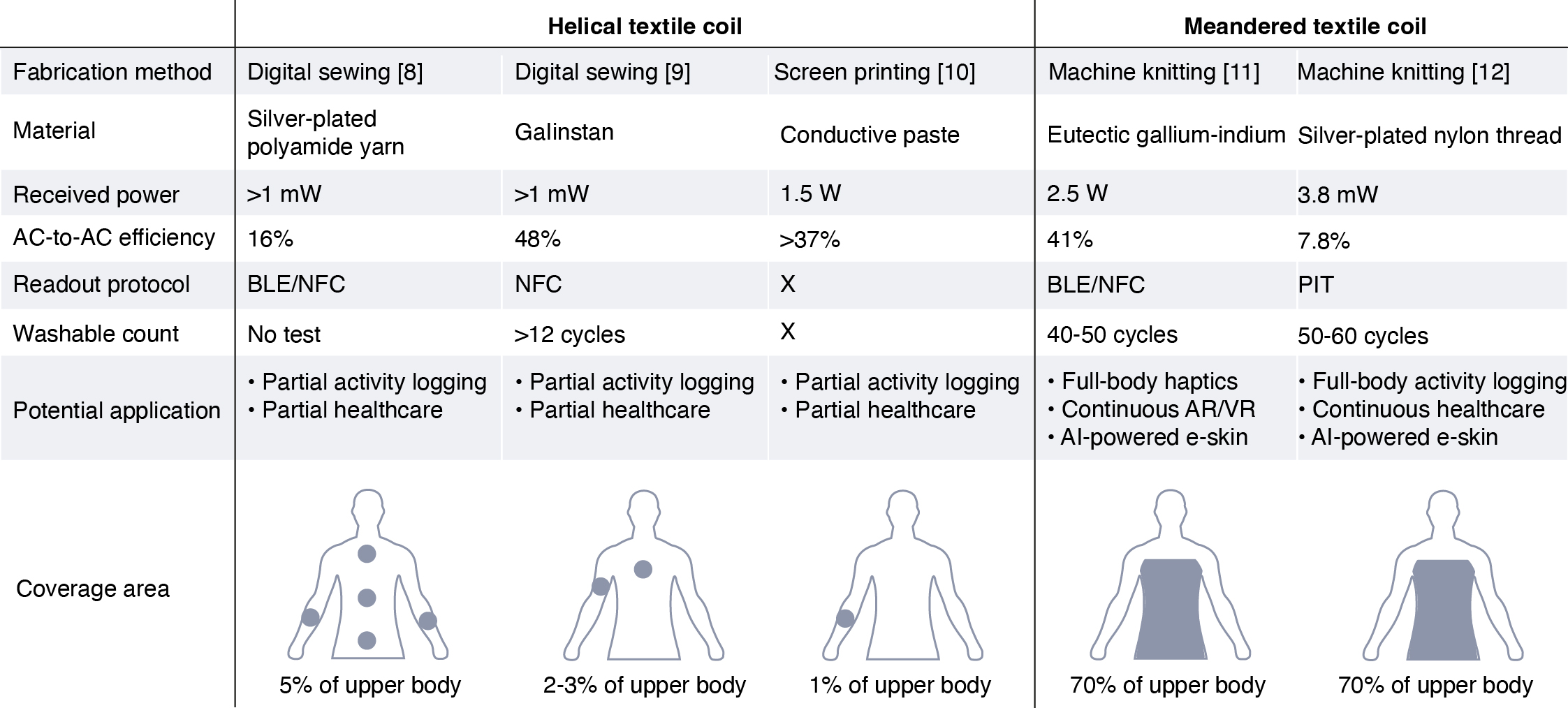}
\label{tab:summary}
\end{table*}

To integrate PIT into the meander coil, the twin meander coil needs to increase the PIT sensitivity. 
This is because increasing the scale of textile coil to whole body weakens the inductive coupling with a small sensor coil, or the passive response.
To solve this challenge, the twin meander coil uses a sensitive readout structure by implementing a pair of two identical meander coils and using them as impedance references for each other.
This twin structure can allows a bridge circuit, which is one of sensitive readout circuit, to achieve a wide-band impedance matching, increasing the PIT sensitivity.
We examined the twin meander coil accurately read out the passive response from a \numproduct{3x3}~\si{\cm} sensor coil occupying $0.3\%$ of the meandered textile area.
The output power from the twin meander coil was decreased to \numrange{10}{100}~\si{\uW}~\cite{Takahashi2022TwinMeadnerCoil}, compared to the output power in NFC protocol~(\textit{i.e.,} \numrange{100}{200}~\si{\mW}).

Finally, we technically compare two types of the meandered textile coils with three types of prior helical textile coils~(see \autoref{tab:summary}).
The conventional helical coil suffers from its narrow-range wireless charging and readout below $5\%$ of the entire body, owing to the EM interaction with the body.
Such a significantly narrow range restricts the device placement to certain parts of the body, which cannot meet the requirement of even the typical application of wearable computing.
By contrast, the meander coil, which confines the inductive field near the body surface, can achieve body-scale wireless charging or readout approximately $70\%$ of the whole body, while ensuring their performance such as \si{\W}-class wireless charging and \si{\uW}-class wireless readout.
Such a large coverage and superior performance can provide continuous operation of multiple wearable devices arbitrarily placed onto the body, in addition to enabling a battery-free, lightweight, compact design of the wearable devices.
For example, meander coil++, which safely transmits \si{\W}-class power to the wearable devices, could remove the heavy, bulky batteries from the haptic devices~\cite{Jung2022Haptics} or AR glasses, providing comfortable haptic perception or immersive AR experience.
In addition, our approach could be compatible with next-generation wearable devices such as artificial-intelligence-powered electronic skin~(AI-powered e-skin)~\cite{Xu2023AIskin} which is preferred to be battery-free for the direct and continuous attachment to the skin.

\section{DISCUSSION}

The meandered textile coil has some limitations.
The fundamental limitation is that the meandered textile coil cannot extend wireless charging and readout range up to naked, non-textile body area such as hand, head, and feet.
To cover such area, one of the promising solutions is to extend a receiver coil connected to the wearable device up to the meandered textile zone.
For example, the meander coil++ delivers power to the neck fan placed to the neck by extending an receiver coil to the area of the meandered textile coil.
Then, the current implementation area is limited to around the torso except for bottoms and sleeves.
This is because the bottoms- or sleeve-based meander coil might inductively interact with the torso-based meander coil, degrading the wireless charging and readout performance.
Time-division-based control scheme using multiple meander coils could mitigate such inductive interference between the coils.
Lastly, the machine knit of the liquid-metal contained in the silicone tube is hard because the tube is so thick for the knitting machine.
One possible solution is to use inlay knitting technique, which can weave a thick tube in the textile during knitting with a customized thread carrier.

\section{CONCLUSION}

This paper introduces the meandered textile coil toward the ubiquitous deployment of continuous, full-body wearable computing.
The localized inductive field of the meander coil allows safe and efficient wireless charging and readout of wearable devices around the whole body.
Moreover, the simple zigzag pattern of the meander coil enables the automatic and fast fabrication of the meander coil using the industrial knitting machine.
With these advancements, the meandered textile coil can integrate the function of continuous full-body wearable computing into everyday clothes.

\section{ACKNOWLEDGMENTS}
This work was supported by JST JPMJAX21K9, JPMJER1501, JPMJMI17F1, JPMJAP2401, and JSPS 22K21343, JP22J11616.

\def\refname{REFERENCES}
\bibliographystyle{IEEEtran}
\bibliography{references}

\begin{thebibliography}{10}
\providecommand{\url}[1]{#1}
\csname url@samestyle\endcsname
\providecommand{\newblock}{\relax}
\providecommand{\bibinfo}[2]{#2}
\providecommand{\BIBentrySTDinterwordspacing}{\spaceskip=0pt\relax}
\providecommand{\BIBentryALTinterwordstretchfactor}{4}
\providecommand{\BIBentryALTinterwordspacing}{\spaceskip=\fontdimen2\font plus
\BIBentryALTinterwordstretchfactor\fontdimen3\font minus \fontdimen4\font\relax}
\providecommand{\BIBforeignlanguage}[2]{{%
\expandafter\ifx\csname l@#1\endcsname\relax
\typeout{** WARNING: IEEEtran.bst: No hyphenation pattern has been}%
\typeout{** loaded for the language `#1'. Using the pattern for}%
\typeout{** the default language instead.}%
\else
\language=\csname l@#1\endcsname
\fi
#2}}
\providecommand{\BIBdecl}{\relax}
\BIBdecl

\bibitem{Sun2023Requirements}
Y.~Sun, Y.-Z. Li, and M.~Yuan, ``Requirements, challenges, and novel ideas for wearables on power supply and energy harvesting,'' \emph{Nano Energy}, vol. 115, p. 108715, 2023.

\bibitem{Wicaksono2020Tailored}
I.~Wicaksono, C.~I. Tucker, T.~Sun, C.~A. Guerrero \emph{et~al.}, ``A tailored, electronic textile conformable suit for large-scale spatiotemporal physiological sensing in vivo,'' \emph{npj Flexible Electronics}, vol.~4, no.~1, p.~5, 2020.

\bibitem{Noda2019I2We}
A.~Noda and H.~Shinoda, ``Inter-ic for wearables (i2we): Power and data transfer over double-sided conductive textile,'' \emph{IEEE Transactions on Biomedical Circuits and Systems}, vol.~13, no.~1, pp. 80--90, 2019.

\bibitem{plug-n-play-eknit2025}
Y.~Li, R.~Takahashi, W.~Yukita, K.~Matsutani \emph{et~al.}, ``Plug-n-play e-knit: prototyping large-area e-textiles using machine-knitted magnetically-repositionable sensor networks,'' in \emph{Proceedings of the Nineteenth International Conference on Tangible, Embedded, and Embodied Interaction}.\hskip 1em plus 0.5em minus 0.4em\relax New York, NY, USA: Association for Computing Machinery, 2025.

\bibitem{Tian2019Metamaterial}
X.~Tian, P.~M. Lee, Y.~J. Tan, T.~L.~Y. Wu \emph{et~al.}, ``Wireless body sensor networks based on metamaterial textiles,'' \emph{Nature Electronics}, vol.~2, no.~6, pp. 243--251, 2019.

\bibitem{Li2021BodyPower}
J.~Li, Y.~Dong, J.~H. Park, and J.~Yoo, ``Body-coupled power transmission and energy harvesting,'' \emph{Nature Electronics}, vol.~4, no.~7, pp. 530--538, 2021.

\bibitem{Lee2020BodyComm}
J.-H. Lee, J.~Ko, K.~Kim, M.~Choi \emph{et~al.}, ``A body channel communication technique utilizing decision feedback equalization,'' \emph{IEEE Access}, vol.~8, pp. 198\,468--198\,481, 2020.

\bibitem{Lin2020TextileCoil}
R.~Lin, H.-J. Kim, S.~Achavananthadith, S.~A. Kurt \emph{et~al.}, ``Wireless battery-free body sensor networks using near-field-enabled clothing,'' \emph{Nature Communications}, vol.~11, no. 444, 2020.

\bibitem{Lin2022GainCoil}
R.~Lin, H.-J. Kim, S.~Achavananthadith, Z.~Xiong \emph{et~al.}, ``Digitally-embroidered liquid metal electronic textiles for wearable wireless systems,'' \emph{Nature Communications}, vol.~13, no. 2190, 2022.

\bibitem{Li2018PrintCoil}
Y.~Li, N.~Grabham, R.~Torah, J.~Tudor \emph{et~al.}, ``Textile-based flexible coils for wireless inductive power transmission,'' \emph{Applied Sciences}, vol.~8, no.~6, 2018.

\bibitem{Takahashi2022MeanderCoil++}
R.~Takahashi, W.~Yukita, T.~Yokota, T.~Someya \emph{et~al.}, ``Meander coil++: A body-scale wireless power transmission using safe-to-body and energy-efficient transmitter coil,'' in \emph{Proceedings of the 2022 CHI Conference on Human Factors in Computing Systems}, 2022.

\bibitem{Takahashi2022TwinMeadnerCoil}
R.~Takahashi, W.~Yukita, T.~Sasatani, T.~Yokota \emph{et~al.}, ``Twin meander coil: Sensitive readout of battery-free on-body wireless sensors using body-scale meander coils,'' \emph{Proc. ACM Interact. Mob. Wearable Ubiquitous Technol.}, vol.~5, no.~4, dec 2022.

\bibitem{sato2025mems}
T.~Sato, S.~Watanabe, R.~Takahashi, W.~Yukita \emph{et~al.}, ``Friction jointing of distributed rigid capacitors to stretchable liquid metal coil for full-body wireless charging clothing,'' in \emph{2025 IEEE 38th International Conference on Micro Electro Mechanical Systems (MEMS)}, 2025, p. 181–184.

\bibitem{Dickey2008eGaIn}
M.~D. Dickey, R.~C. Chiechi, R.~J. Larsen, E.~A. Weiss \emph{et~al.}, ``Eutectic gallium-indium (egain): A liquid metal alloy for the formation of stable structures in microchannels at room temperature,'' \emph{Advanced Functional Materials}, vol.~18, no.~7, pp. 1097--1104, 2008.

\bibitem{ICNIRP2020}
\BIBentryALTinterwordspacing
(2020) Icnirp guidelines. [Online]. Available: \url{https://www.icnirp.org/cms/upload/publications/ICNIRPrfgdl2020.pdf}
\BIBentrySTDinterwordspacing

\bibitem{Takahashi2020TelemetRing}
R.~Takahashi, M.~Fukumoto, C.~Han, T.~Sasatani \emph{et~al.}, ``Telemetring: A batteryless and wireless ring-shaped keyboard using passive inductive telemetry,'' in \emph{Proceedings of the 33rd Annual ACM Symposium on User Interface Software and Technology}, 2020, p. 1161–1168.

\bibitem{picoRing2024takahashi}
R.~Takahashi, E.~Whitmire, R.~Boldu, S.~Ng \emph{et~al.}, ``picoring: battery-free rings for subtle thumb-to-index input,'' in \emph{Proceedings of the 37th Annual ACM Symposium on User Interface Software and Technology}, 2024.

\bibitem{Jung2022Haptics}
Y.~H. Jung, J.-Y. Yoo, A.~V{\'a}zquez-Guardado, J.-H. Kim \emph{et~al.}, ``A wireless haptic interface for programmable patterns of touch across large areas of the skin,'' \emph{Nature Electronics}, vol.~5, no.~6, pp. 374--385, 2022.

\bibitem{Xu2023AIskin}
C.~Xu, S.~A. Solomon, and W.~Gao, ``Artificial intelligence-powered electronic skin,'' \emph{Nature Machine Intelligence}, vol.~5, no.~12, pp. 1344--1355, 2023.

\end{thebibliography}

\begin{IEEEbiography}
{Ryo Takahashi}{\,} is currently a project assistant professor in the Graduate School of Engineering at the University of Tokyo, The University of Tokyo, Tokyo, Japan. 
His research interests lie broadly in wearable computing, electric textiles, and battery-free, wireless sensor networks.
He received the Ph.D. degree in Graduate School of Engineering from The University of Tokyo in 2023. 
He is a member of IEEE, ACM, IEICE, and IPSJ.
Contact him at takahashi@akg.t.u-tokyo.ac.jp.
\end{IEEEbiography}

\begin{IEEEbiography}
{Yoshihiro Kawahara}{\,} is currently a professor with the Department of Electrical Engineering and Information Systems, Graduate School of Engineering, The University of Tokyo, Tokyo, Japan. 
He joined the faculty in 2005. 
His research interests include the areas of computer networks and ubiquitous and mobile computing. 
He received the Ph.D. degree in information communication engineering in 2005. 
He is a member of IEEE, IEICE, and IPSJ. 
Contact him at kawahara@akg.t.u-tokyo.ac.jp.
\end{IEEEbiography}

\end{document}